\documentclass[aps,prl,twocolumn,superscriptaddress]{revtex4-1}

\usepackage{amssymb}
\usepackage{amsmath}
\usepackage{amsfonts}
\usepackage{graphicx}
\usepackage{color}
\usepackage{xspace}
\usepackage{ulem}
\usepackage{mathtools}
\usepackage{hhline}

\newcommand{\AddrEdin}{
 Tait Institute, School of Physics and Astronomy, University of Edinburgh, Edinburgh, EH9 3JZ, United Kingdom
}

\newcommand{\AddrCP}{Schmid College of Science, Chapman University, Orange, CA 92866, USA}

\newcommand{\AddrVandy}{Department of Physics and Astronomy, Vanderbilt University, Nashville, TN 37235, USA}

\newcommand{\AddrDortmund}{
Fakult\"at f\"ur Physik, Technische Universit\"at Dortmund, 44221 Dortmund, Germany
}

\newcommand{\AddrAve}{Departamento de F\'{\i}sica da Universidade de Aveiro and CIDMA,  Campus de Santiago, 3810-183 Aveiro, Portugal
}

\begin{document}


\title{Knotty inflation and the dimensionality of spacetime}

\author{Arjun Berera}    \email{ab@ph.ed.ac.uk}\affiliation{\AddrEdin} 
\author{Roman V. Buniy}    \email{buniy@chapman.edu}\affiliation{\AddrCP}
\author{Thomas W. Kephart}    \email{thomas.w.kephart@vanderbilt.edu}\affiliation{\AddrVandy}
\author{Heinrich P\"as}    \email{heinrich.paes@tu-dortmund.de}\affiliation{\AddrDortmund}
\author{Jo\~{a}o G.~Rosa\footnote{Also at Departamento de F\'{\i}sica e Astronomia, Faculdade de Ci\^encias da Universidade do Porto, Rua do Campo Alegre 687, 4169-007 Porto, Portugal.}}    \email{joao.rosa@ua.pt}\affiliation{\AddrAve}

\date{\today}

\begin{abstract}
We suggest a structure for the vacuum comprised of a network of
tightly knotted/linked flux tubes formed in a QCD-like cosmological
phase transition and show that such a network can drive cosmological
inflation.  As the network can be topologically stable only in three space dimensions, this scenario provides a dynamical explanation for the existence of exactly three large spatial dimensions in our Universe.
\end{abstract}

\maketitle

{\it Introduction:}
Although the  question of why our Universe has exactly three (large) spatial dimensions is one of the most profound puzzles in 
cosmology - especially in view of quantum gravity scenarios such as string theory which assume 9 or 10 space dimensions at the fundamental level - it is actually only occasionally addressed in the literature \cite{3D}. In this paper we propose a topological explanation for the dimensionality of space-time based on the idea that inflation is driven by a tightly knotted network of flux tubes generated in a cosmological phase transition
and the fact that one-dimensional knots are topologically stable only in exactly three space dimensions ("Knotty Inflation").

The main idea of the model is connected to the fact that, in non-Abelian gauge theories analogous to QCD, chromoelectric flux tends to become confined and the resulting tube-like structures can be treated as effective one-dimensional objects or strings. The formation of flux tubes is behind the basic models of hadronization in QCD, with flux strings connecting  quark-antiquark pairs giving rise to a linear potential that confines them into mesons and prevents the existence of free colored charges. A similar description with more complicated flux tube shapes also describes the formation of baryons and glueballs, the latter denoting bound states of pure flux.

During the phase transition from a generic quark-gluon plasma to a hadron gas, a large number of flux tubes may fill up the whole Universe and form an intricate network of both open and closed flux tubes, with some similarities to what is known as the `spaghetti vacuum' \cite{Nielsen:1978rm}, potentially with a large linking and/or crossings number density provided the flux tube density is sufficiently large. A related idea has been suggested recently in \cite{bjorken}. It has been shown that such a network is stabilized by a topological conservation law \cite{Bekenstein:1992ut}.

This network will then tend to relax into a tight configuration that fills the whole of space in order that each tube minimizes its energy, similarly to the description of glueballs in terms of knotted/linked flux tubes proposed in \cite{Buniy:2002yx}. The network then decouples from the Hubble flow. 


The energy stored in this tight network will thus provide an effective cosmological constant which is homogeneous on scales large compared to the typical flux tube width and on average isotropic if the flux tubes are randomly oriented. This effective cosmological constant will thus trigger a period of inflation in the early universe that lasts until the network decays through reconnection, tube breaking and other quantum effects. Network breaking will proceed until a gas of hadrons and radiation is formed and the standard cosmological evolution begins.

In higher-dimensional theories, such as string theory, gauge theories are typically confined to hypersurfaces of $p$ spatial dimensions known as D$p$-branes, to which open strings with gauge, fermionic and scalar degrees of freedom are attached. These fundamental strings behave like point-like objects on energy scales smaller than the fundamental string scale, $M_s$, and so, for a confinement scale $\Lambda<M_s$, we nevertheless expect the formation of tube-like structures connecting color charges in higher-dimensional gauge theories. The tube cross sections will have a higher-dimensional geometry, in the simplest case with $S^{p-2}$ topology, but will in any case behave effectively as one-dimensional objects.

The main difference between the dynamics of confinement in $3+1$ and higher dimensional gauge theories lies in the fact that flux tube knots and links will only be topologically stable (or metastable) in three spatial dimensions, being otherwise able to unknot/unlink in the extra-dimensions. Given that this is an essential feature for the formation of a tight network of flux that decouples from the Hubble flow and provides an effective cosmological constant, confinement can only lead to inflation for gauge theories in D3-branes. 

Hence, considering a generic universe in a higher-dimensional space with gauge theories living on branes of different dimensionalities, only D3-branes will inflate with the formation of a tight flux tube network during a cosmological phase transition. This can thus explain why our universe has only three large spatial dimensions in the absence of other sources of vacuum energy.

It is the purpose of this Letter to outline the main features of this mechanism and discuss the necessary properties that a flux tube network must exhibit in order to yield a successful inflationary model in three dimensions. While leaving a detailed modeling of the network's evolution and observational predictions of this scenario for a future work, we will describe the most important physical processes that drive the network's formation, inflationary dynamics and subsequent decay, giving quantitative estimates for the relevant time and energy scales.


{\it Flux tubes in gauge theories:}
Understanding confinement in QCD and gauge theories in general is one of the most important problems in particle physics. Due to the strong coupling and non-perturbative nature of confinement, this is an intrinsically hard problem that can currently only be studied accurately in the context of numerical simulations on the lattice. It is nevertheless widely accepted that the confinement of chromoelectric charges, such as quarks and anti-quarks in QCD, is associated with a squeezing of chromoelectric flux lines into tube-like structures connecting the charges. These flux tubes then give rise to a potential energy that grows linearly with the distance between pairs of quarks and anti-quarks, therefore leading to their confinement in mesons, baryons and potentially other hadronic structures. Flux tubes can be described e.g.~in terms of Abrikosov-Nielsen-Olesen vortices \cite{Nielsen:1973cs, Abrikosov:1956sx}  in dual superconductor models of confinement (see \cite{Ripka:2003vv} for a review).

Reference \cite{Buniy:2002yx} pointed out that closed flux tubes may have non-trivial topologies, including knotted tubes and links between distinct closed flux tubes. The degree of ``knottedness" of each configuration is then associated with a conserved topological charge, analogous e.g.~to the (Abelian) magnetic helicity that characterizes conducting fluids in magnetohydrodynamics \cite{MHD}. 
Any such topologically non-trivial flux tube configuration will necessarily relax into an equilibrium state by minimizing its length and therefore its energy. These equilibrium states then correspond to the tightest knotted or linked configurations with a given topological charge and number of flux {\it quanta}.
This thus motivates us to consider the description employed in \cite{Buniy:2002yx}, where the chromoelectric field $F_{0i}$ (where $i=1,2,3$ in three spatial dimensions) is confined in knotted/linked tube-like structures that carry one quantum of chromoelectric flux. The relevant (static) Lagrangian density is given by:
\begin{equation} \label{Lagrangian}
\mathcal{L}={1\over2}\mathrm{tr}\left[F_{0i} F^{0i}\right]+\mathrm{tr}\lambda \left[ {\Phi_E\over (\pi a^2)}- n^iF_{0i}\right]~, 
\end{equation}
up to the addition of a constant energy density as in the MIT bag model \cite{MIT}, where the last term enforces flux conservation across the  tube sections through a Lagrange multiplier $\lambda$, with $n^i$ denoting the unit vector normal to the tube's cross section. We expect the flux tube radius $a$ to be given by the confinement scale $\Lambda$, such that $a\sim \Lambda^{-1}$ in natural units, which we will employ in our discussion henceforth. This yields the equations of motion for the gauge field:
\begin{equation} \label{gauge_eom}
D^0 (F_{0i}-\lambda n_i)=0~, \qquad D^{i}(F_{0i}-\lambda n_i)=0~, 
\end{equation}
the solution of which is given by a constant chromoelectric field
\begin{equation} \label{chromoelectric_field}
F_{0i}={\Phi_E\over \pi a^2}n_i
\end{equation}
which vanishes outside the flux tube. The energy density inside each flux tube is then given by:
\begin{equation} \label{energy_density}
\rho_E={1\over2}{\mathrm{tr}\Phi_E^2\over (\pi a^2)^2}\sim {\Lambda^4\over 2\pi^2}~,
\end{equation}
up to $\mathcal{O}(1)$ factors that will not affect our discussion.

{\it Network formation:} Let us then hypothesize that there exists a confining gauge theory, analogous to QCD but with a generic high-energy confinement scale $\Lambda$, and which lives on a three-dimensional brane within a higher-dimensional compact space. For simplicity, we will denote the particles charged under this gauge group as quarks and gluons, although one must bear in mind that these are not the known QCD fields but rather novel degrees of freedom that must decouple from the low-energy effective theory.

In the early Universe, for temperatures above a certain critical value $T_c\sim \Lambda$, these quarks and gluons are essentially free, forming a plasma that we will assume is close to thermal equilibrium and dominates the energy balance in the Universe. This plasma has an energy density $\rho_R (T)=(\pi^2/30) g_* T^4$, where $g_*$ denotes the number of relativistic degrees of freedom, which includes at least the massless gluons and the light standard model fields of the theory. Due to expansion, the temperature of the plasma will decrease until it becomes lower than the critical value, at which point all the free gluons become confined in both open and closed flux tubes, the former connecting quark-antiquark pairs. Given that
\begin{equation} \label{energy_comparison}
\rho_R (T_c)\sim g_*\Lambda^4 \sim \rho_E~,
\end{equation}
we expect that, at the confining phase transition, a large number of flux tubes is formed within each Hubble volume. Such a large density will naturally lead to a large number of crossings between the flux tubes and, hence, to a large number of knots and links of different configurations. This is analogous to the behavior observed when a string is tumbled inside a box of fixed volume, where the probability of knotting increases very quickly above a certain critical string length and which effectively corresponds to a critical density \cite{Raymer}. The flux tube network will then be endowed with a non-trivial topology, which can be metastable in three spatial dimensions and that,  as we describe below, can be sufficiently long-lived to drive a period of inflation in the early Universe.


{\it Inflation:} After the phase transition, Hubble expansion will tend to stretch the flux tubes, while maintaining their radius (set by the strong non-perturbative dynamics), and increase the size of the gaps in between them. However, as argued above, the presence of knots and links makes the network behave differently from a system of isolated flux tubes, endowing it with a rigidity and making it try to relax into a tight equilibrium configuration. Flux tubes will then shrink and approach each other in order to maximize the fraction of the spatial volume that they can occupy given the topological constraints. In addition, the number of knots and links in the network will decrease due to the decay processes that we describe below. These will make the network progressively less tight, with the tightest configuration occupying an increasingly smaller fraction of the spatial volume.

We then expect a network to remain tight if:
\begin{equation} \label{time_hierarchy}
\tau_{tight} \ll H^{-1} \ll \tau_{decay}~, 
\end{equation}
i.e.~if it relaxes into a tight configuration more quickly than expansion and knots/links are stable on the Hubble time scale. If the network is sufficiently dense at the phase transition, i.e.~if the gaps between flux tubes are not much wider than the tube radius, we expect $\tau_{tight}\lesssim \Lambda^{-1}$, and the network can in principle relax into a tight configuration within a Hubble time for $\Lambda \gg H$. In this case, the tube width is also much smaller than the Hubble radius, so that the network is homogeneous on near-horizon and super-horizon scales. It should also be, on average, isotropic if quarks and gluons are randomly distributed in the thermal plasma at the transition.

Let us see that a tight network will behave as an effective cosmological constant as long as the above time scale hierarchy is satisfied. The ``tightness" of the network can be evaluated in terms of the fraction of the spatial volume occupied by the flux tubes, $f$, which will in general depend on the space and time coordinates. On average, we expect a network that is homogeneous and isotropic on scales larger than the tube radius, such that $f=f(t)$ at leading order. The network thus has an average energy density $\rho=f \rho_E$, with the first law of thermodynamics yielding:
\begin{equation} \label{first_law}
d(\rho V)= {df\over dV}(\rho_E V) dV+f\rho_E dV= -pdV~,
\end{equation}
neglecting decay processes and consequent heat transfer. From this we conclude that the network has an effective pressure: 
\begin{equation} \label{pressure}
p=-\left(1+{1\over3}{d\log f\over dN_e}\right)\rho~,
\end{equation}
 where $N_e$ denotes the number of e-folds of expansion, and hence an equation of state parameter $w=p/\rho \simeq -1$ for $\dot{f}/f \ll 3H$.

The network will initially relax into a tight configuration with $d\log f/dN_e > 0$, yielding a {\it phantom}-like equation of state $w<-1$ \cite{Caldwell:1999ew}. This phantom behavior is often considered in dynamical dark energy models, as well as in dark matter scenarios with bulk viscosity \cite{Disconzi:2014oda}, and it would be interesting to consider this behavior in the context e.g.~of axionic strings \cite{Vilenkin:1982ks}. 

The network will then subsequently remain close to a tight configuration, with a fixed or at most slowly varying filling fraction $f$. The rigidity of the network, which results from the presence of a large density of knots and links, will thus decouple it from the Hubble flow and generate an effective cosmological constant.

As the number of knots and links decreases due to network decay, $f$ will then decrease such that $w\gtrsim -1$, thus yielding a period of inflation with an equation of state analogous to that of a canonical slowly-rolling scalar field.  Note that, in the absence of the topological charge that maintains the tightness of the network, the filling fraction would decrease quickly and no accelerated expansion could be obtained. 

The Friedmann equation yields, during inflation,
\begin{equation} \label{Friedmann}
H^2\simeq {f \Lambda^4\over 6\pi^2 M_P^2}~,
\end{equation}
which implies that $\Lambda < H$ for a sub-Planckian confinement scale, with e.g~a phase transition at the grand unification scale, $\Lambda \sim 10^{16}$ GeV, yielding $H/\Lambda\sim 10^{-3}$ for $f\lesssim 1$. This shows that the flux tube network will in general be on average homogeneous and isotropic on super-horizon scales and that relaxation may in principle occur in less than a Hubble time, as required above.


{\it Network decay and reheating}: The flux tube network is metastable in three spatial dimensions, since its degree of knotting/linking is reduced by string breaking and other reconnection events. 

Two flux tubes may unknot or unlink by quantum tunneling. We can estimate this by treating the intersection between two flux tubes as a (non-relativistic) particle of mass $m\sim \rho_E a^3\sim \Lambda/2\pi^2$ that tunnels through a potential barrier of length $\sim a$ and height $\Delta V\gtrsim \Lambda$. The standard tunneling rate computation then yields a typical lifetime:
\begin{equation} \label{tunneling}
\tau_t H={1\over \beta}(aH) \exp\left({1\over \pi} \sqrt{\Delta V\over \Lambda}\right)~,
\end{equation}
where $\beta$ is the typical flux tube velocity. The tunneling rate is thus suppressed for a sufficiently large potential barrier separating the flux tubes. Such a barrier has been obtained for SU(2) flux tubes within the dual superconductor description of confinement \cite{Baker:1986dp} but, as far as we are aware, more generic computations have yet to be developed. In bag models, the flux enclosed within a tube is supported by surface currents and one can expect the repulsion between currents in adjacent tube sections to effectively provide a `hard' boundary for the flux tubes, making tunneling more difficult.

If tunneling is indeed suppressed, flux tubes will primarily break, and hence unknot or unlink, through the well-known Schwinger effect \cite{Schwinger:1951nm}, where $Q\bar{Q}$ pairs are produced in the approximately constant chromoelectric field enclosed within each flux tube. For $N_f$ quark flavors of mass $m_Q$, using that the string tension $\kappa= \pi a^2\rho_E\sim \Lambda^2/(2\pi)$, we obtain for the characteristic lifetime of a flux tube of length $l$: 
\begin{equation} \label{string_breaking}
\tau_b H\sim  {8\pi^4\over N_f}(aH)^2(lH)^{-1} \exp\left(2\pi {m_Q^2\over \Lambda^2}\right)~.
\end{equation}
This implies that the lifetime for string breaking can be parametrically large if there are no light quarks in the spectrum of the gauge theory, $m_Q\gg \Lambda$, even for $a\ll H^{-1}$ and flux tubes as long as the Hubble radius. Thus, the confining gauge theory responsible for a sufficiently long period of inflation in the proposed scenario has to be distinct from QCD, since in the latter case knots and links between flux tubes will quickly decay through the production of up and down quark-antiquark pairs.

Given the exponential suppression of the string breaking and tunneling rates, it is thus not difficult to envisage scenarios where the network remains tight for the 50-60 e-folds of accelerated expansion required to explain the present flatness and homogeneity of the Universe.

String breaking will eventually lead to a system of unknotted/unlinked flux tubes, and thus a gas of hadrons. These may be unstable and decay into radiation, including in principle the Standard Model particles. The details of this process depend, of course, on how the confining gauge theory at a high energy scale $\Lambda$ is related to the low-energy physics, but it is clear that the proposed model naturally includes a mechanism for reheating the Universe after inflation. It is also possible for a significant amount of radiation to be produced during the inflationary period, if e.g.~small loops unlink from the network and decay while the network is still tight, thus potentially leading to a warm inflation model  \cite{wi}.

{\it Cosmological perturbations}: The flux tube network is, on average, homogeneous on scales larger than the tube radius, which as shown above can be parametrically below the Hubble radius. Nevertheless, fluctuations in the energy density of the flux tube network may arise on super-horizon scales and become imprinted on the curvature of space-time, thus seeding the temperature anisotropies in the Cosmic Microwave Background and the Large Scale Structure of the present Universe. 

Curvature perturbations in a perfect fluid driving a period of accelerated expansion propagate with an imaginary sound speed $c_s^2= dp/d\rho = w <-1/3$, which would make the system unstable to small fluctuations as opposed to the canonical scalar field models of inflation. However, the flux tube network has a non-vanishing rigidity, as a result of the presence of numerous knots and links. The network thus behaves more like an elastic solid than a perfect fluid, similarly to the networks of topological defects considered in \cite{Battye:2005hw, Battye:2005ik}. The anisotropic stress inherent to elastic solids allows for the propagation of both longitudinal and transverse waves and the associated sound speed $c_s^2= w + (4/3)\mu/(\rho+p) $ can be real for a sufficiently large shear modulus $\mu$  \cite{Sitwell:2013kza}.

Inflationary models with elastic solids may lead to a nearly scale-invariant spectrum of both curvature and tensor perturbations, with interesting differences from scalar field models \cite{Sitwell:2013kza}. Firstly, despite the absence of non-adiabatic modes, perturbations evolve on super-horizon scales due to the presence of anisotropic stress, although maintaining the relevant scaling between different super-horizon modes. The overall amplitude of the spectrum thus depends on the details of reheating, in our case of the decay of the flux tube network. Anisotropic stress will also act as a source for tensor perturbations, potentially modifying the primordial gravitational waves spectrum with respect to canonical models.
 
  Although concrete observational predictions require a detailed modeling of the properties and dynamics of the flux tube network, which is beyond the scope of this Letter, it is worth mentioning that a red-tilted curvature spectrum can only be obtained in elastic solid models of inflation for a slowly varying equation of state. We expect this to be the case in our model since a knotted network will be close, but not exactly in, a tight configuration, as a result of the opposing effects of Hubble expansion, relaxation and unknotting events.
 
 {\it Conclusion:}
A knotted/linked flux tube network formed in a QCD-like phase transition can provide a natural source of inflation. Furthermore, this may explain why we live in three large spatial dimensions, since knotted/linked one-dimensional tubes are topologically unstable in higher-dimensional space-times. 
We realize that the picture developed in this paper
may also be applied to a model of dark energy, which would
eliminate the need for an ultra-light scalar field.

The volume filling fraction $f$ is not a fully dynamical variable because it does not enter the field theory Lagrangian in the above calculations.  One can easily (and most likely without any unpleasant consequences) remedy this drawback by considering a fully dynamical model.  As a concrete, simple and non-trivial example of such, we can take the abelian Nielsen-Olesen model, where the role of the fraction $f$ is played by time-dependent characteristics of spatial extents of the scalar and gauge fields.  It is plausible that main dynamical characteristics of flux tube knotting and the evolution of a knotted network in terms of tightening, Hubble expansion and decay in this abelian model will not change significantly when a generic non-abelian gauge theory is used instead.  Although exact solutions in such models are unavailable, appropriate approximations should be enough to establish quantitative features to support the rest of our calculations.  We will present the detailed analysis based on the abelian Nielsen-Olesen  model elsewhere \cite{BBKPR}.

\begin{acknowledgments}
{\bf Acknowledgments}

AB is supported by STFC. The work of TWK was supported by US DOE grant DE-SC0010504. HP acknowledges kind hospitality and support at Vanderbilt University and by the Alexander von Humboldt-Foundation. JGR is supported by the FCT grant SFRH/BPD/85969/2012 and partially by the grant PTDC/FIS/116625/2010, the CIDMA strategic project UID/MAT/04106/2013 and the Marie Curie action NRHEP-295189-FP7-PEOPLE-2011-IRSES. Some of this work was developed at the Isaac Newton Institute and we thank them for their kind hospitality.
\end{acknowledgments}

\end{document}